\documentclass[10 pt,final,journal,letterpaper,oneside,twocolumn]{IEEEtran}
\usepackage{multicol}   
\usepackage{multirow}
\usepackage[pdftex]{graphicx}
\usepackage[ruled]{algorithm2e}
\usepackage{algorithmic}
\usepackage[small]{caption}
\usepackage{float}
\usepackage{amssymb,amsmath}
\usepackage{graphicx}
\usepackage{subfigure}

\usepackage{xspace}
\usepackage{fancyhdr}
\usepackage{amssymb,amsmath}
\usepackage{textcomp} 
\usepackage{epstopdf}
\usepackage{bbding}
\usepackage{cite} 
\usepackage{color}
\usepackage{pifont}
\usepackage{flushend}
\usepackage{hyperref}
\usepackage{cleveref}
\begin{document}

\title{\huge Joint Optimization for Secure Ambient Backscatter Communication in NOMA-enabled IoT Networks}
\author{Wali Ullah Khan, Furqan Jameel, Asim Ihsan, Omer Waqar, and Manzoor Ahmed\thanks{Wali Ullah Khan is with Interdisciplinary Centre for Security, Reliability and Trust (SnT), University of Luxembourg, 1855 Luxembourg City, Luxembourg (emails: waliullah.khan@uni.lu, waliullahkhan30@gmail.com). 

Furqan Jameel is with the Department of Communications and Networking, Aalto University, 02150 Espoo, Finland (email: furqanjameel01@gmail.com). 

Asim Ihsan is with the Department of Communication and Information Engineering, Shanghai Jiao Tong University, Shanghai 200240, China (email: ihsanasim@sjtu.edu.cn). 

Omer Waqar is with the Department of Engineering, Thompson Rivers University (TRU), British Columbia (BC), Canada (email: owaqar@tru.ca). 

Manzoor Ahmed is with the College of Computer Science and
Technology, Qingdao University, Qingdao 266071, China (email: manzoor.achakzai@gmail.com).
}\vspace{-0.5cm}}%

\markboth{}%
{Shell \MakeLowercase{\textit{et al.}}: Bare Demo of IEEEtran.cls for IEEE Journals}

\maketitle

\begin{abstract}
Non-orthogonal multiple access (NOMA) has emerged as a novel air interface technology for massive connectivity in sixth-generation (6G) era. The recent integration of NOMA in backscatter communication (BC) has triggered significant research interest due to its applications in low-powered Internet of Things (IoT) networks. However, the link security aspect of these networks has not been well investigated. This article provides a new optimization framework for improving the physical layer security of the NOMA ambient BC system. Our system model takes into account the simultaneous operation of NOMA IoT users and the backscatter node (BN) in the presence of multiple eavesdroppers (EDs). The EDs in the surrounding area can overhear the communication of base station (BS) and BN due to the wireless broadcast transmission. Thus, the main objective is to enhance the link security by optimizing the BN reflection coefficient and BS transmit power. To gauge the performance of the proposed scheme, we also present the suboptimal NOMA and conventional orthogonal multiple access as benchmark schemes. Monte Carlo simulation results demonstrate the superiority of the NOMA BC scheme over the pure NOMA scheme without BC and conventional orthogonal multiple access scheme in terms of system secrecy rate.
\end{abstract}


\IEEEpeerreviewmaketitle

\section{Introduction}
Beyond fifth-generation (5G) communication technologies are expected to provide high-spectrum and energy-efficient systems along with lower transmission latency \cite{9261140}. As an important member of multiple access family, non-orthogonal multiple access (NOMA) has the capability of connecting multiple Internet of Things (IoT) users using the same spectrum and time resources \cite{li2020uav}. It can be achieved by applying superposition coding and successive interference cancellation (SIC) techniques \cite{khan2018efficient1}. 

On the parallel avenue, backscatter communications (BC) has received an overwhelming research interest due to the ultra-low-power nature of transmission \cite{ji2020joint}. BC allows backscatter node (BN) to modulate and retransmit data symbols using the incident radio frequency (RF) signal. To operate the circuit, BN harvests a fractional of energy from the RF signal \cite{jameeltime}. Due to the available RF source, the BN does not need oscillators to generate carrier signals \cite{jameel2019simultaneous}.

Besides the features of
BC systems, there are also some limitations
of BC systems. One of the main limitations is that BN requires a dedicated RF source for data transmission \cite{li2021physical}. Although these systems have been developed in radio frequency identification (RFID) tags and deployed in grocery stores and libraries, it may not be suitable for energy-constrained IoT networks \cite{liu2018optimal}. To overcome this limitation, a system called ambient BC has emerged as a new paradigm for low-powered IoT networks. The ambient BC utilizes the incident RF signals of WiFi, TV, BS to perform data transmission in a battery-free manner. This feature significantly reduces the deployment cost of ambient BC in large-scale IoT networks. \cite{han2017wirelessly}. However, the research of BC is still in the early stage due to the novelty of these systems. Recently, NOMA has been proposed as a new radio access technology for BC in ultra low-powered IoT networks. \cite{zhang2019backscatter}. Although the integration of NOMA in BC is a hot topic for researchers, the existing works have mostly studied the performance evaluation and throughput maximization of the systems.
\subsection{Related Work, Motivation and Contributions}
The rapid evolution of NOMA has lead to the point where researchers have started to explore its applications in BC. For instance, by integrating NOMA in BC, the research of \cite{zhang2019backscatter} has studied the outage performance and ergodic capacity of the system. In \cite{guo2018design}, the authors have proposed a BC with NOMA in both fading and non-fading systems. The work of \cite{farajzadeh2019uav} has optimized trajectory and altitude of unmanned aerial vehicle (UAV) to reduce the energy consumption of BC system. Le {\em et al}. \cite{8847703} have studied the outage performance of NOMA and BC in multi-antenna system. Reference \cite{8851217} has explored the optimization framework of time and reflection coefficient for BC to enhance the spectral efficiency of the system. Khan {\em et al.} \cite{khan2021backscatterV2X} have also investigated the performance of backscatter communication in vehicle-to-everything networks. 

Besides, in \cite{zeb2019backscatter}, Zeb {\em et al.} have also proposed a hybrid BC to investigate the spectral efficiency and outage of the NOMA system. The authors of \cite{8761990,li2020secrecy}, have also discussed the outage and intercept probabilities of BC systems. Reference \cite{liao2020resource} has proposed a joint optimization framework of subcarrier assignment, time allocation and reflection coefficient to maximize the throughput of NOMA symbiotic radio system. In another study \cite{li2019secure}, a symbiotic radio communication has designed to enhancement the outage secrecy rate (SR) of MISO system. Moreover, researchers have also investigated the performance of BC using learning techniques \cite{9024401,jameel2020reinforcement}. Of late, the problems of spectral efficiency have also explored for NOMA BC systems \cite{khanjoint,9261963,khan2021backscatter}.

Although a considerable amount of works has been done on both NOMA and BC. However, a joint optimization framework to improve the SR of the system in the presence of multiple eavesdroppers (EDs) has not yet been investigated, to the best of our knowledge. Thus, the main objective of this article is to provide an optimization framework for maximizing the SR of NOMA ambient BC under multiple non-colluding EDs. More specifically, we jointly optimize the reflection coefficient of BN and transmit power of base station (BS) subject to the reflection power of BN and power budget of BS. The problem of SR maximization is formulated non-convex followed by an efficient solution obtained by dual theory, where dual variables are iteratively calculated. The important contributions of this article can be summarized as follow:
\begin{itemize}
\item A NOMA ambient BC system is considered, where a BS transmits a superimposed signal to its associated IoT users in downlink direction. Meanwhile, the BN and multiple EDs also receive the BS superimposed signal, where the EDs try to decode and overhear the information of IoT users. The BN also uses the same signal to add its own data and reflect it to the IoT users in uplink direction. Thus, the objective of this work is to enhance the SR of the BC system by jointly optimizing the reflection coefficient of BN transmit power of BS.
\item The mathematical problem of SR maximization is formulated as non-convex, where the constraints of a maximum reflection coefficient at BN and transmit power at BS are taken into account. The joint solution is obtained through Lagrangian dual method, where the Lagrangian variables are iteratively updated. Moreover, pure NOMA optimization without BC and conventional OMA BC system are also provided as the benchmark schemes.
\item To validate the closed-form mathematical solutions, extensive results based on Monte Carlo simulations are also provided. The simulation results demonstrate that the proposed joint optimization scheme significantly outperforms the suboptimal NOMA and optimal OMA schemes. The results show that NOMA system with BC significantly improve the system SR compared to those of pure NOMA without BC and conventional OMA BC systems. 
\end{itemize}
The remainder of this paper is structured as follows: Section II discusses different consideration and assumptions of system model followed by optimization problem formulation. Section III presents the joint solutions. Section IV presents and discusses the numerical simulation results, while the concluding remarks and future works are provided in Section V.

\section{System Model and Joint Problem Formulation}
\subsection{System Model}
\begin{figure*}
\centering
\includegraphics[width=0.60\textwidth]{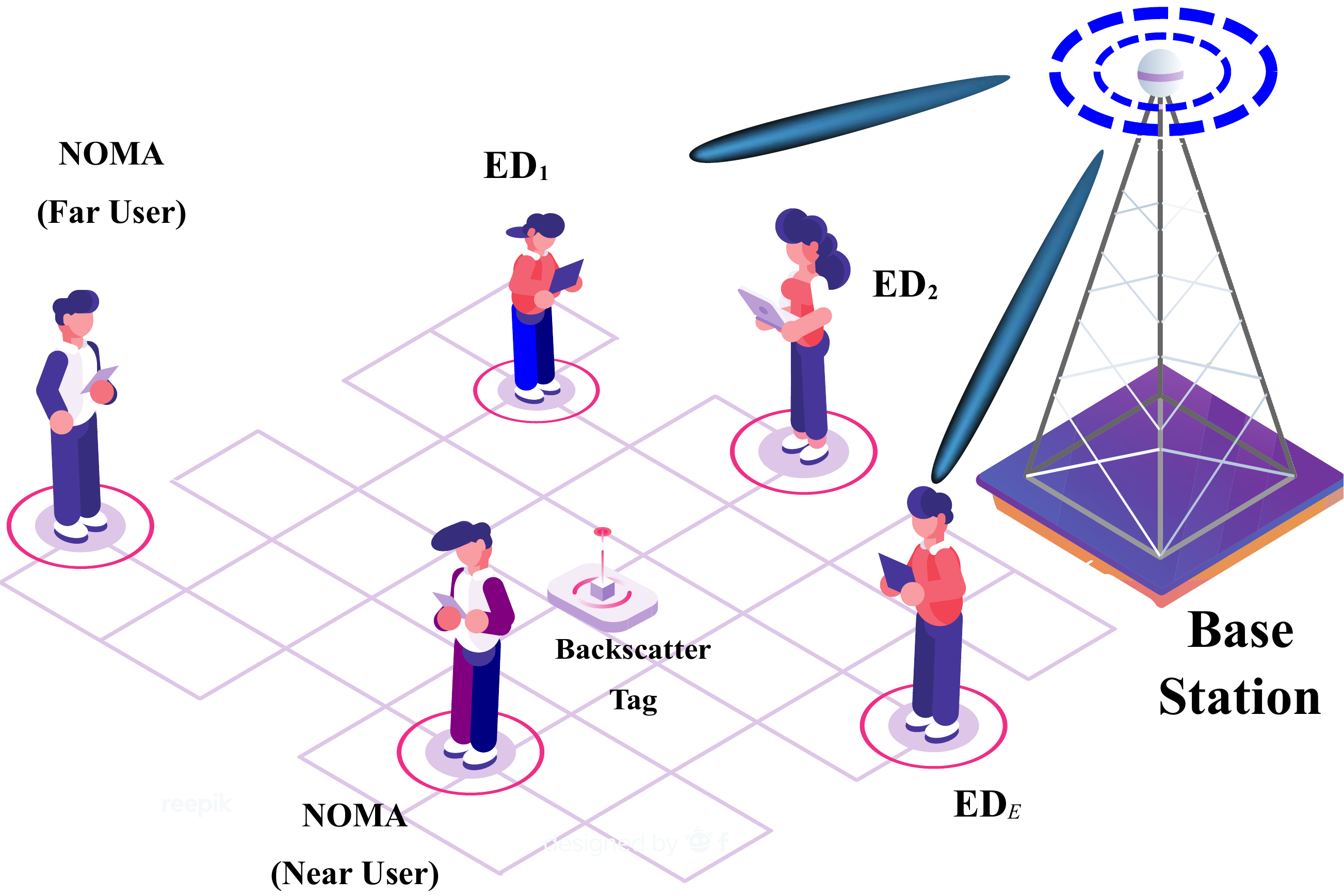}
\caption{System model of NOMA ambient BC.}
\label{block}
\end{figure*}

Let us consider a downlink BC network, where a BS communicates with two IoT users utilizing power-domain NOMA, as shown in Fig. \ref{block}. It is assumed that IoT user $U_n$ is located near to the BS and has good channel conditions while another IoT user (denoted as $U_f$) is located away from the BS and has the weak channel conditions. During the NOMA transmission, a BN and multiple EDs also receive the superimposed signal from BS. Due to the broadcast nature of the backscattered signal, it is most vulnerable to eavesdropping attacks. Therefore, we consider a worst-case scenario in which the EDs aim to decode the signals from both the BS and reflected signals from BN. Let $\mathcal E=\{E_k|k=1,2,\dots,K\}$ be the set of EDs. The BN uses the incident signal from BS, modulates data, and retransmit it to $U_n$. Thus, BN is intended to communicate with $U_n$ due to a very low-power budget, whereby $U_n$ can also act as a reader. 
Following the recent literature on BC and NOMA, channel information and signal synchronization are available. Moreover, the processing delay and the impact of power consumption at BN can be safely ignored. According to the NOMA \cite{khan2020efficient}, $U_n$ applies SIC and decode the signal of $U_f$. After this, $U_n$ decodes its own signal, and then it decodes the signal from the BN in the end \cite{li2019secure}. However, the far user $U_f$ cannot apply SIC and decode its signal by treating the signals from BN and $U_n$ a noise.
\subsubsection{Received Signal at $U_n$}

The superimposed signal of BS can be written as
\begin{align}
z=\sqrt{p\omega}s_n+\sqrt{p(1-\omega)}s_f,
\end{align} 
where $s_n$ and $s_f$ are the unit-power desired signals of $U_n$ and $U_f$, respectively. Moreover, $p$ is the total power of BS and $0<\omega < 0.5$ denotes the power allocation coefficient. According to the power-domain NOMA, the BS assigns more power to $U_f$ compared to $U_n$ to guarantee successful SIC at receivers and to ensure fairness between both IoT users \cite{8861078}. As we mentioned above, the BN utilizes the incident superimposed signal from BS to modulate its signal $x(t)$ such that $\mathbb E[|x(t)|^2]=1$, and retransmit it to $U_n$. Thus, the received signal at $U_n$ can be represented as
%%%%%%%%%%%%%%%%%%%%%%%%%%%%%%%%%%%%%%%%%%
\begin{align}
y_n=G_nz+\sqrt{\alpha}G_bzH_nx(t)+\beta_n,
\end{align}
%%%%%%%%%%%%%%%%%%%%%%%%%%%%%%%%%%%%%%%%%%
where $G_b$, $G_n$ and $H_n$ represent the channel coefficients from the BS to BN, from the BS to $U_n$, and from BN to $U_n$, respectively. The term $\alpha$ is the reflection power of BN and $\beta_n$ is the zero mean additive white Gaussian noise (AWGN) with variance $\sigma^2$.

Due to the double fading effect, the data signal of BN to $U_n$ is much weaker than the direct signal from BS to $U_n$. Thus, $U_n$ applies SIC to decode and remove its own signal as well as the signal of $U_f$ from BS before decoding its desired signal from BN. Thereby, the signal to interference plus noise ratio (SINR) at $U_n$ to decode the signal of $U_f$ is given by
\begin{align}
\gamma_{n,f}=\frac{|G_n|^2p(1-\omega)}{|G_n|^2p\omega+|H_n|^2\alpha|G_b|^2(p\omega+p(1-\omega))+\sigma^2},
\end{align}   
After removing $s_f$, $U_n$ decodes its own signal. The SINR at $U_n$ to decode its own signal can be written as
\begin{align}
\gamma_{n,n}=\frac{|G_n|^2p\omega}{|H_n|^2\alpha|G_b|^2(p\omega+p(1-\omega))+\sigma^2},
\end{align}

Finally, $U_n$ will decode the data signal of BN. Therefore, the SINR at $U_n$ for decoding BN signal can be stated as
\begin{align}
\gamma_{n,b}=\frac{|H_n|^2\alpha|G_b|^2(p\omega+p(1-\omega))}{\sigma^2}.
\end{align}
%%%%%%%%%%%%%%%%%%%%%%%%%%%%%%%%%%%%%%%%%%
\subsubsection{Received Signal at $U_f$}
The received signal of $U_f$ can be expressed as
%%%%%%%%%%%%%%%%%%%%%%%%%%%%%%%%%%%%%%%%%%
\begin{align}
y_f=G_fz+\sqrt{\alpha}G_bzH_fx(t)+\beta_f,
\end{align}
%%%%%%%%%%%%%%%%%%%%%%%%%%%%%%%%%%%%%%%%%%
where $G_f$ and $H_f$ denote the channel coefficients from BS to $U_f$ and from BN to $U_f$, respectively. The term $\beta_f$ stands for zero mean AWGN with variance $\sigma^2$. Please note that $U_f$ decodes its own signal without SIC. Therefore, the SINR at $U_f$ for decoding its desired signal can be stated as
%%%%%%%%%%%%%%%%%%%%%%%%%%%%%%%%%%%%%%%%%%
\begin{align}
\gamma_{f,f}=\frac{|G_f|^2p(1-\omega)}{|G_f|^2p\omega+|H_f|^2\alpha|G_b|^2(p\omega+p(1-\omega))+\sigma^2}.
\end{align} 
\subsubsection{Received Signal at $E_k$}
The EDs also receive the superimposed signal from BS and reflected signal from BN. Therefore, the signal that $E_k$ receives can be derived as
%%%%%%%%%%%%%%%%%%%%%%%%%%%%%%%%%%%%%%%%%%
\begin{align}
y_b=G_kz+\sqrt{\alpha}G_bzH_kx(t)+\beta_k,
\end{align}
%%%%%%%%%%%%%%%%%%%%%%%%%%%%%%%%%%%%%%%%%%
where $H_k$ and $G_k$ are the channel coefficients from BN to $E_k$ and from BS to $E_k$, respectively.
The SINR at $E_k$ to decode the signal from BN can be formulated as
\begin{align}
\gamma_{k,b}=\frac{|H_k|^2\alpha|G_b|^2(p\omega+p(1-\omega))}{|G_k|^2(p\omega+p(1-\omega))+\sigma^2},
\end{align}  
\subsection{Secrecy Rate Maximization Problem}
In this work, we optimize SR\footnote{Generally, it is difficult to calculate the exact capacity of BC \cite{qian2018iot}. Similar to \cite{lyu2017optimal}, we assume the maximum achievable capacity of BC.} against multiple passive EDs, i.e., non-colluding. Thus, the SR of $U_n$ from BN in the presence of multiple EDs can be written as 
\begin{align}
R_{sec}=[\log_2(1+\gamma_{n,b})-\log_2(1+\underset{{k}}{\text{max}}(\gamma_{k,b}))]^+.
\end{align} 
%%%%%%%%%%%%%%%%%%%%%%%%%%%%%%%%%%%%%%%%%%
%%%%%%%%%%%%%%%%%%%%%%%%%%%%%%%%%%%%%%%%%%
It is worth mentioning that the weakest signal among all is the signal received from BN. Due to this reason, it is more prone to eavesdropping attacks. Thus, the main objective of this paper is to increase the SR of backscattered message to $U_n$ through optimizing the reflection coefficient of BN and transmit power of BS. This optimization framework is subject to the maximum reflection power of BN and BS transmit power, as per NOMA protocol\footnote{Due to the double fading effect of BC \cite{zhang2019backscatter}, the interference at $U_f$ from BN brings a negligible change in the performance and can be safely ignored.}. Thus, the following secrecy optimization problem can be written:
%%%%%%%%%%%%%%%%%%%%%%%%%%%
\begin{alignat}{2}
\text{OP}\quad& \underset{{\alpha,\omega}}{\text{maximize}}\ R_{sec}\label{11}\\
\quad\text{s.t.}\quad & 0\leq\alpha\leq1,\tag{11a}\label{11a}\\
& p\omega\leq p(1-\omega),\tag{11b}\label{11b}
\end{alignat}
%%%%%%%%%%%%%%%%%%%%%%%%%% 
where (\ref{11}) represents the objective function of SR maximization. Constraint in (\ref{11a}) limits the reflection coefficient of BN between 0 and 1 while constraint in (\ref{11b}) is the power allocation according to the NOMA principle. 
\section{Proposed Solution}

The SR maximization problem in OP is non-convex optimization because of the coupled variables and interference term, therefore, we adopt Lagrangian dual method to obtain the efficient solutions. This is the low complex iterative method for solving optimization problems. Let us derive the dual-problem of OP as
%%%%%%%%%%%%%%%%%%%%%%%%%% 
\begin{align}
\underset{{\zeta\geq0,\lambda\geq0}}{\text{min}}\ \underset{{\alpha\geq0,\omega\geq0}}{\text{max}}\ R_{sec}-\zeta(\alpha-1)-\lambda(p\omega-p(1-\omega)), 
\end{align}
%%%%%%%%%%%%%%%%%%%%%%%%%% 
where $\zeta$ and $\lambda$ are the associated dual variables. Removing the constant terms and substitute the value of $R_{sec}$, the problem can be rewritten as
%%%%%%%%%%%%%%%%%%%%%%%%%% 
\begin{align}
&\underset{{\zeta\geq0,\lambda\geq0}}{\text{min}}\ \underset{{\alpha\geq0,\omega\geq0}}{\text{max}} \log_2\bigg(1+\frac{|H_n|^2\alpha|G_b|^2(p\omega+p(1-\omega))}{\sigma^2}\bigg)\nonumber\\&-\log_2\bigg(1+\frac{|H_k|^2\alpha|G_b|^2(p\omega+p(1-\omega))}{|G_k|^2(p\omega+p(1-\omega))+\sigma^2}\bigg)-\zeta\alpha-\lambda p\omega,\label{13} 
\end{align}
%%%%%%%%%%%%%%%%%%%%%%%%%% 
Exploiting Karush-Kuhn-Tucker (KKT) conditions to the internal maximization and calculate the closed-form expression of $\alpha$ and $\omega$, we obtain
%%%%%%%%%%%%%%%%%%%%%%%%%% 
\begin{align}
\alpha^*=\bigg(\frac{-\Pi_2\pm\sqrt{\Pi_2^2-4\Pi_1(|G_b|^2p\sigma^2+p\Pi_3/\zeta)}}{2\Pi_1}\bigg)^+,\label{14}
\end{align} 
%%%%%%%%%%%%%%%%%%%%%%%%%% 
and 
%%%%%%%%%%%%%%%%%%%%%%%%%% 
\begin{align}
\omega^*=\bigg(\frac{(|H_k|^2+1)|H_n|^2\alpha^2|G_b|^4p^3|H_k|^2|G_k|^2}{\lambda p}\bigg)^+,\label{15}
\end{align} 
%%%%%%%%%%%%%%%%%%%%%%%%%% 
where $[\eta]^+=\text{max}(0,\eta)$, $\Pi_1=|G_b|^4p^2|H_n|^2|H_k|^2$, $\Pi_2=|H_b|^4p^2|H_n|^2+|H_n|^2|G_b|^2p\sigma^2+|H_k|^2|G_b|^2p\sigma^2$, and $\Pi_3=|H_k|^2|H_b|^2\sigma^2-|H_n|^2|G_b|^4p-|H_n|^2|G_b|^2\sigma^2$, respectively. With optimal $\alpha^*$ and $\omega^*$, the dual problem in (\ref{13}) becomes
%%%%%%%%%%%%%%%%%%%%%%%%%% 
\begin{align}
&\underset{{\zeta\geq0,\lambda\geq0}}{\text{min}} \log_2\bigg(1+\frac{|H_n|^2\alpha^*|G_b|^2(p\omega^*+p(1-\omega^*))}{\sigma^2}\bigg)\nonumber\\&-\log_2\bigg(1+\frac{|H_k|^2\alpha^*|G_b|^2(p\omega^*+p(1-\omega^*))}{|G_k|^2(p\omega^*+p(1-\omega^*))+\sigma^2}\bigg)\nonumber\\&-\zeta\alpha^*-\lambda p\omega^*,\label{16} 
\end{align}
%%%%%%%%%%%%%%%%%%%%%%%%%% 
Finally, the dual variables can be iteratively calculated as \cite{khan2020multiobj}
%%%%%%%%%%%%%%%%%%%%%%%%%%
\begin{align}
\zeta(1+j)=\zeta(j)+\delta(j)(\alpha^*-1),\label{Grad1}
\end{align}
%%%%%%%%%%%%%%%%%%%%%%%%%%
%%%%%%%%%%%%%%%%%%%%%%%%%%
\begin{align}
\lambda(1+j)=\lambda(j)+\delta(j)(p\omega^*-p(1-\omega^*)),\label{Grad2}
\end{align}
%%%%%%%%%%%%%%%%%%%%%%%%%%
where $j$ indexes the iteration number and $\delta\geq0$ is the positive step size. In each iteration of (\ref{Grad1}) and (\ref{Grad2}), the efficient reflection coefficient of BN and optimal transmit power
at BS are obtained using (\ref{14}) and (\ref{15}). At convergence, the optimal values of dual variables as well as
of reflection coefficient and transmit power are obtained.

It is worth-mentioning to discuss the complexity of the BC system using Lagrangian dual method. In this work, the complexity is calculated in terms of iteration needed for the convergence. It can be observed that the complexity of the proposed BC system depends on (17) and (18). If the convergence precision is denoted as $\Theta$ and the number of total iterations required for convergence can be stated as $J$, then the complexity of the proposed optimization framework can be expressed as $\mathcal O(\frac{1}{\Theta}\log(J))$. The proposed iterative process of the Lagrangian dual method is also shown in Algorithm 1.
%%%%%%%%%%%%%%%%%%%%%%%%%%%%%%%%%%%%%%%%%%
   \begin{algorithm}[t]
  Initialize all the system parameters, i.e., power budget of BS, number of EDs, reflection coefficient of BN, noise variance, and set iteration number as $j=0$.       
    
    \While{not converge}{Calculate $\alpha$ and $\omega$ using (14) and (15).\\Iteratively update $\zeta$ and $\lambda$ using (17) and (18).
        }
    Return $\alpha^*$, $\omega*$.
    \caption{Proposed Lagrangian dual method.}
   \end{algorithm}
   %%%%%%%%%%%%%%%%%%%%%%%%%%%%%%%%%%%%%%%%%%
   
\section{Numerical Results and Discussion}
Here, we aim to provide key findings and relevant analysis as a result of 10000 Monte Carlo simulations. Unless stated otherwise, the system parameters for performing simulation are set as follow: the channels among various devices in the system are assumed to follow Rayleigh fading, such that $G_n\sim\mathcal{CN}(0,\theta_n)$, $G_f\sim\mathcal{CN}(0,\theta_f)$, $G_b\sim\mathcal{CN}(0,\theta_b)$, $G_k\sim\mathcal{CN}(0,\theta_k)$, $H_n\sim\mathcal{CN}(0,\theta_{H,n})$, $H_f\sim\mathcal{CN}(0,\theta_{H,f})$, and $H_k\sim\mathcal{CN}(0,\theta_{H,k})$, where $\theta_i=0.1$ with $i\in\{n,f,b,k\}$. In addition, the maximum number of EDs are $K=10$, the total power budget at BS is set as $p=10$ W, and the maximum reflection coefficient of BN is set as $0\leq\alpha\leq1$. We compare the proposed joint optimization scheme (stated as Optimal NOMA AmBackCom) with suboptimal scheme (denoted as SubOptimal NOMA AmBackCom) and conventional OMA scheme (represented as Optimal OMA AmBackCom). Note that SubOptimal NOMA AmBackCom refers to optimization scheme with fixed power allocation at BS and optimal reflection coefficient of BN.
%%%%%%%%%%%%%%%%
\begin{figure}[!t]
\centering
\includegraphics[width=0.45\textwidth]{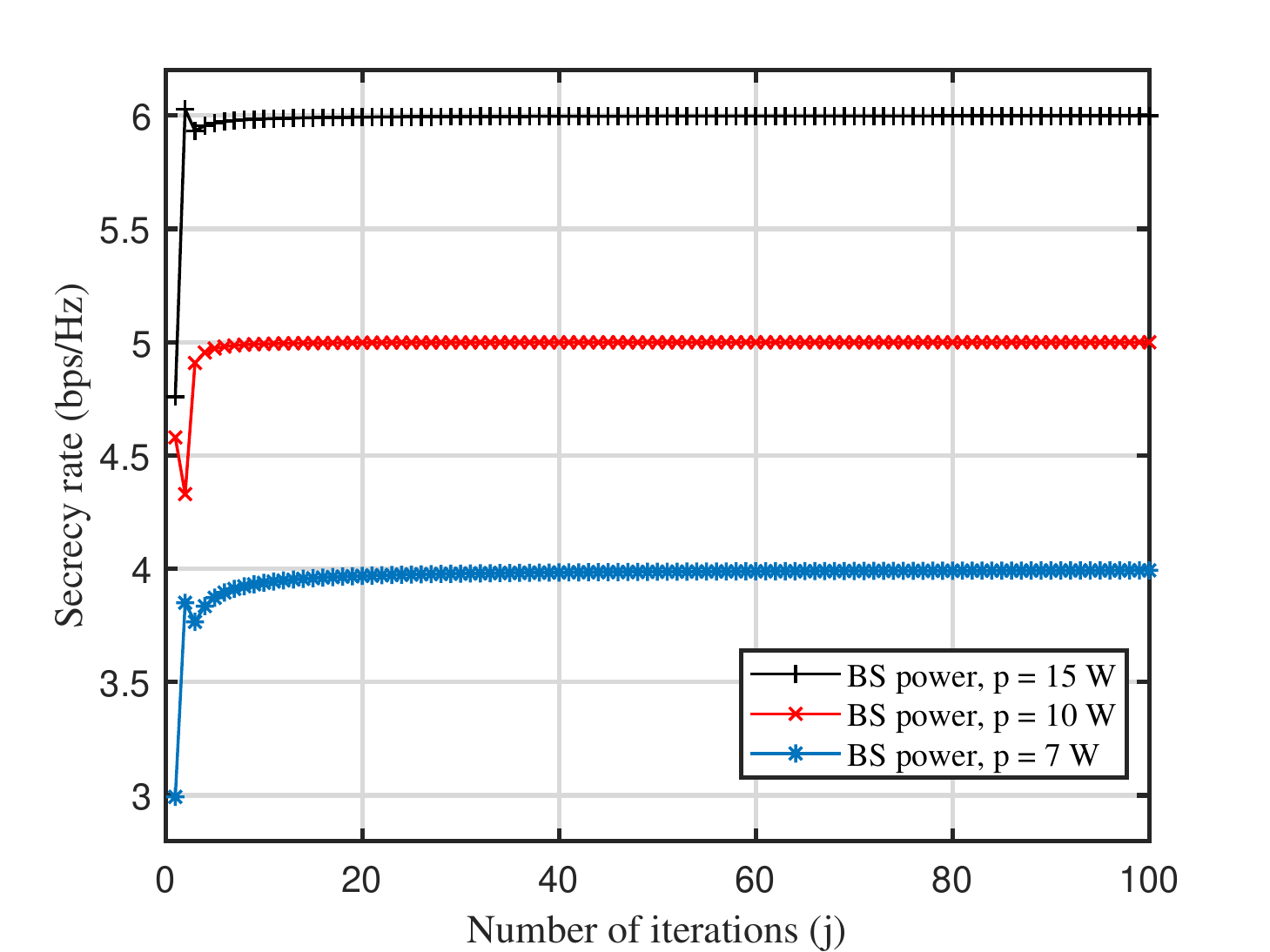}
\caption{SR versus the number of iterations with $K=5$.}
\label{SR-Conv}
\end{figure}
%%%%%%%%%%%%%%%%

\subsection{Eavesdropping under OMA}
We provide a detailed comparison of the proposed Optimal NOMA AmBackCom and SubOptimal NOMA AmBackCom schemes with the conventional Optimal OMA AmBackCom scheme. In this work, we use time-division multiple-access as an OMA scheme. According to this scheme, $U_n$ and $U_f$ can receive their desired messages in separate time intervals. More specifically, the BS transmits to only one user at one time. Here, in order to avoid interference, the BN uses the second time slot to reflect the data to $U_n$. Moreover, no backscattering takes place when the BS transmits data to $U_n$. 
%%%%%%%%%%%%%%%%
\begin{figure}[!t]
\centering
\includegraphics[width=0.45\textwidth]{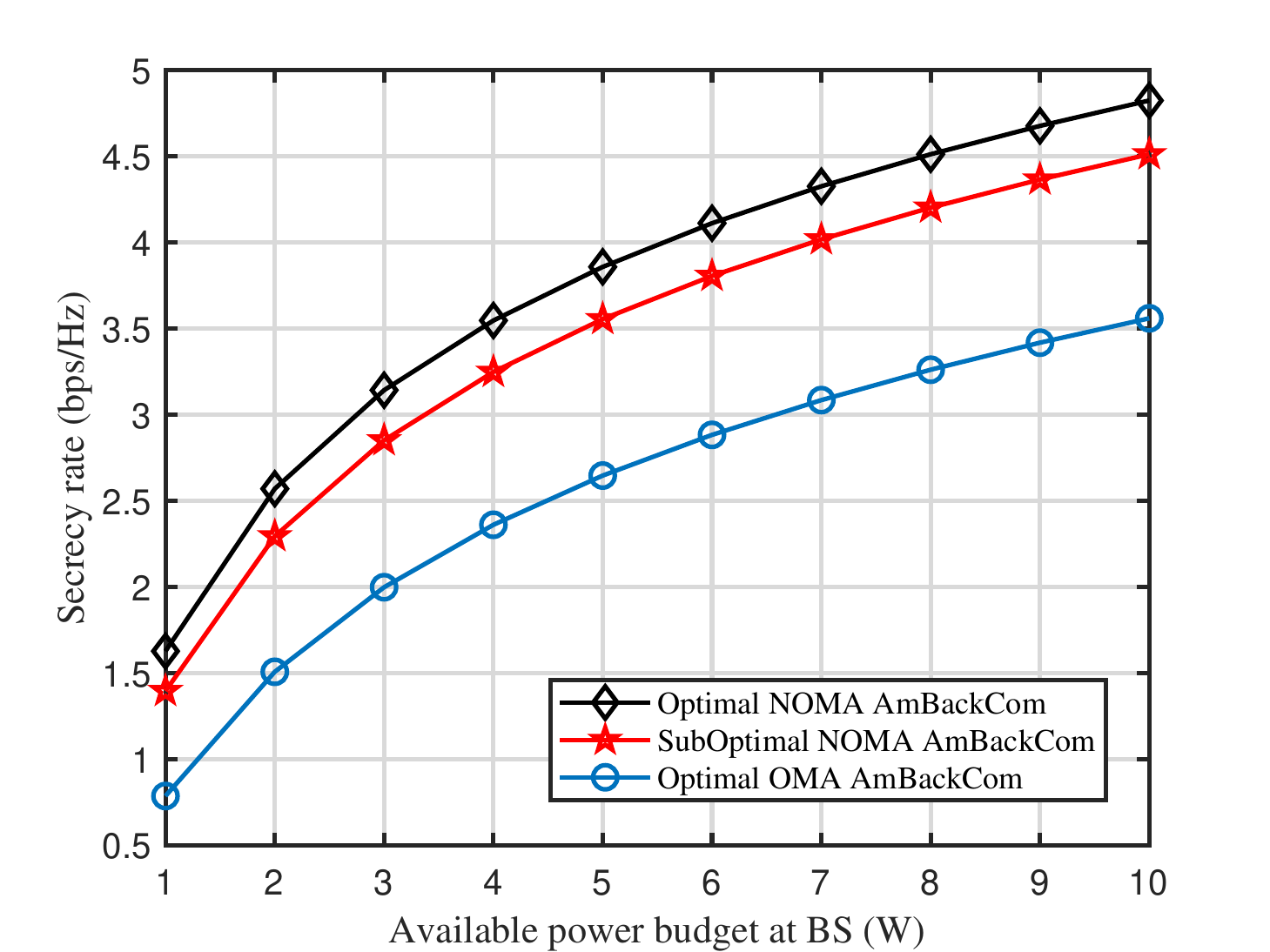}
\caption{SR versus source power ($K=5$).}
\label{SRvsSP}
\end{figure}
%%%%%%%%%%%%%%%%
\begin{figure}[!t]
\centering
\includegraphics[width=0.45\textwidth]{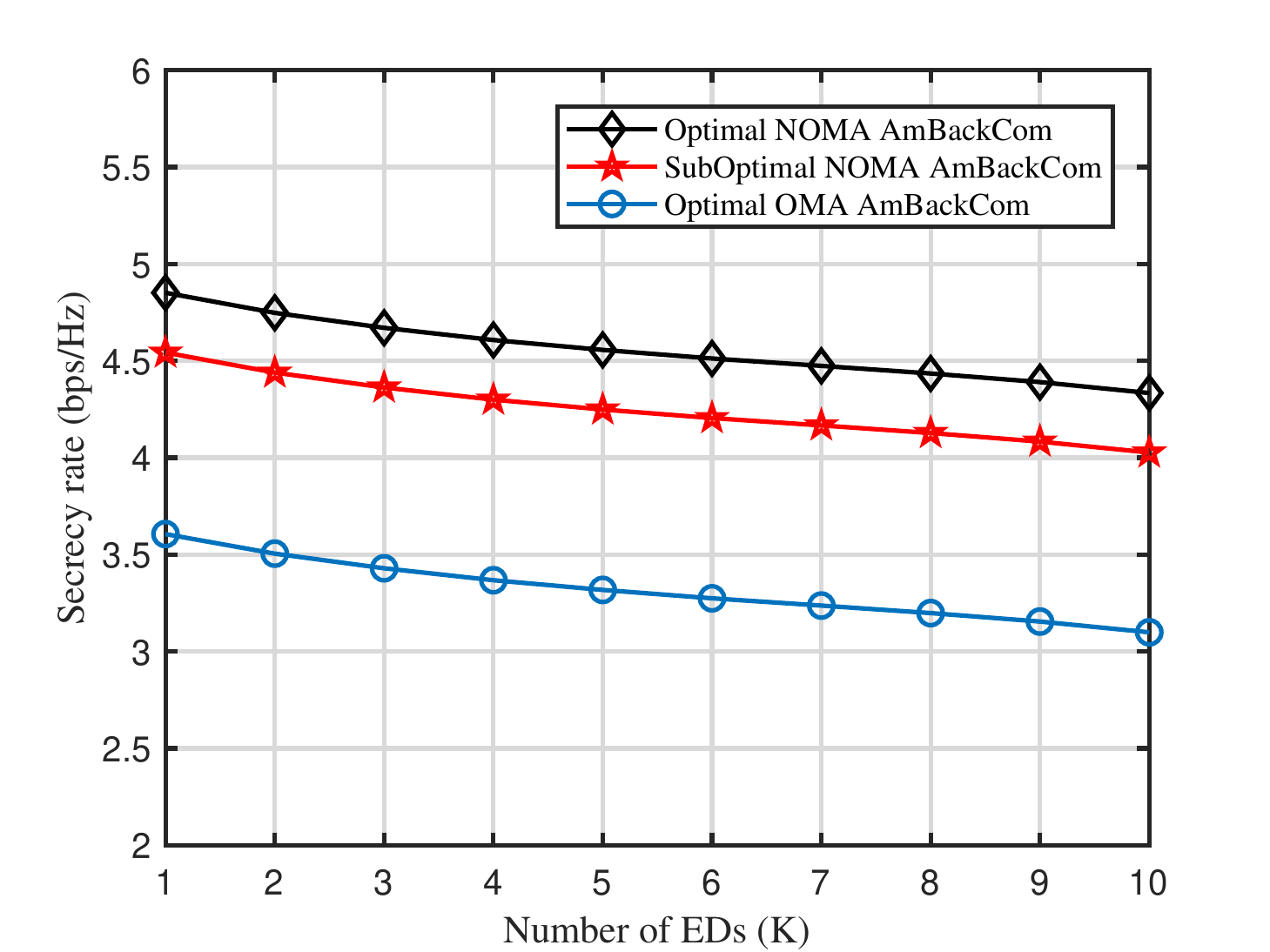}
\caption{SR versus the number of EDs when $p=10$ W.}
\label{SRvsED}
\end{figure}
\subsection{Comparative Analysis and Discussions}
Fig. \ref{SR-Conv} illustrates the convergence of the proposed optimization framework with a different power budget of BS, i.e., $p=7,10,15$ W. It can be evident that the proposed optimization scheme converges within 20 iterations, which show the low complexity nature of the algorithm. Moreover, the proposed method with a larger power budget at BS can achieve a high SR than a lower power budget.

Fig. \ref{SRvsSP} depicts the performance of the proposed framework for increasing the values of BS transmit power. It can be seen that at lower transmit power values, the curves of SR for the proposed Optimal NOMA AmBackCom scheme, SubOptimal NOMA AmBackCom scheme, and Optimal OMA AmBackCom scheme are close to each other. However, as the transmit power of BS increases, the SR improves, and the gap between the curves of the proposed Optimal NOMA AmBackCom scheme and the other benchmark schemes also increases. Moreover, it can also be noted that our proposed NOMA AmBackCom scheme performs significantly better than the conventional OMA AmBackCom scheme.

To show the effectiveness of our proposed framework, Fig. \ref{SRvsED} provides the SR of the proposed Optimal NOMA AmBackCom scheme and the other two benchmark schemes against the increasing number of EDs. It can be seen that the SR decreases when the number of EDs in the system increases. This is because a large number improves the ED's capability of decoding IoT user's information. However, the proposed Optimal NOMA AmBackCom and SubOptimal NOMA AmBackCom schemes maintain the stable SR, despite increasing EDs. In contrast, the secrecy performance of Optimal OMA AmBackCom significantly drops when the total number of EDs in the system approaches 10.
%%%%%%%%%%%%%%%%
\begin{figure}[!t]
\centering
\includegraphics[width=0.45\textwidth]{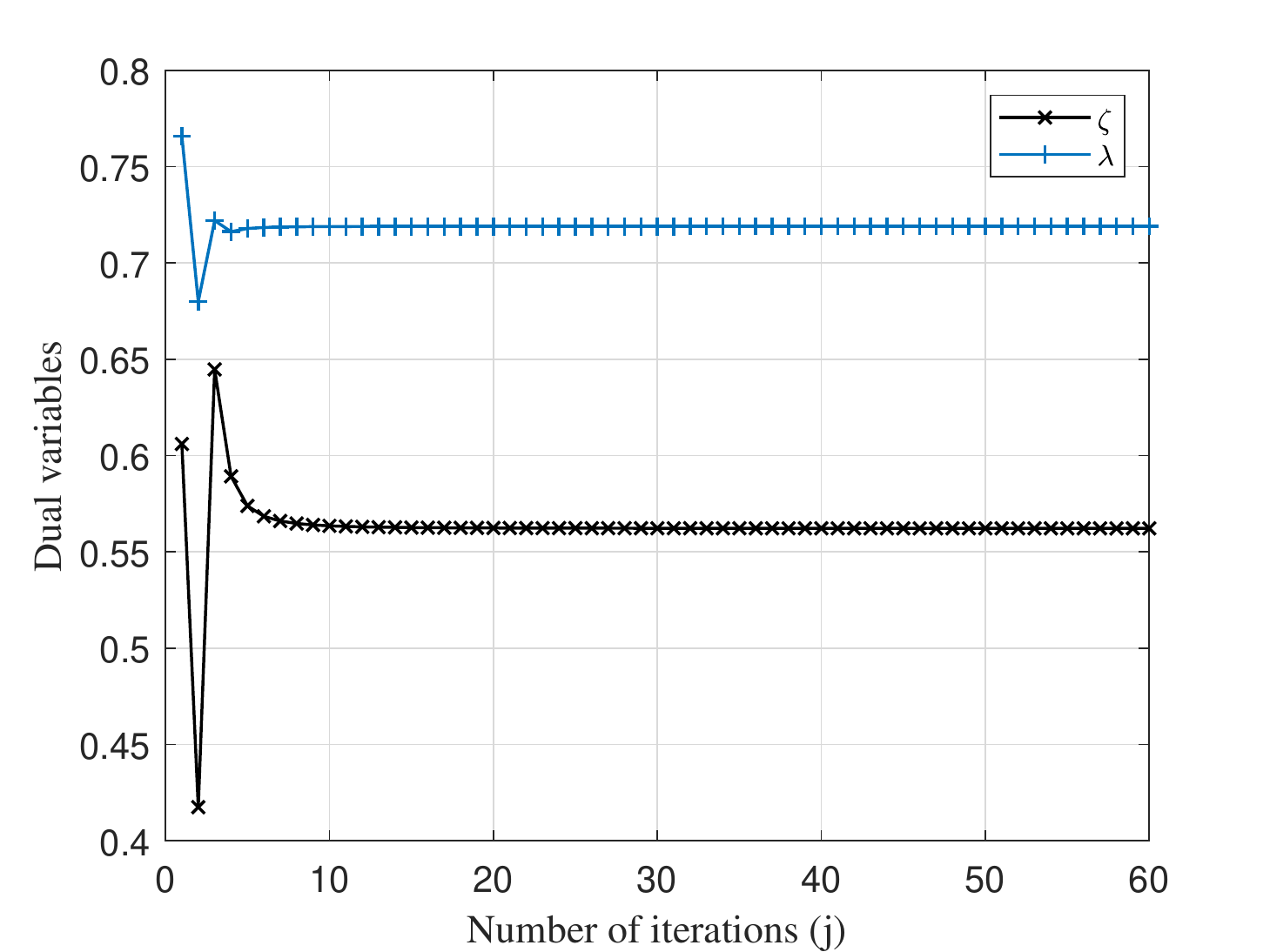}
\caption{Convergence of dual variables $\zeta$ and $\lambda$.}
\label{Conv}
\end{figure}

Last but not least, Fig. \ref{Conv} plots the number of iterations required for convergence of the proposed Optimal NOMA AmBackCom scheme versus the dual variables involve in the solution. It can be seen from the result that $\zeta$ and $\lambda$ converge around 20 iterations. It shows that the Lagrangian dual optimization method not only improves the link security of the network but also ensures very low complexity. 

\section{Conclusion}
Ambient BC and NOMA are expected to pave the way for realizing dense and ultra low-powered IoT networks. This article provided a joint optimization framework for guaranteeing the physical layer security of these massive networks. In particular, we have presented a SR maximization solution to provide physical layer security against multiple EDs in the system. The performance of the proposed optimal scheme for multiple EDs was compared with suboptimal NOMA and OMA BC schemes. Our work can be extended to the case where multiple EDs in the system are cooperating (colluding) with each other. Moreover, the proposed framework can also be extended to a multi-cell NOMA BC system where mutual interference exists among various cells. These attractive yet explored problems will be investigated in the future study.

\bibliographystyle{IEEEtran}
\bibliography{egbib}

\end{document}